\documentstyle[editedvolume]{crckapb10} 

\begin{opening}
\title{COSMIC STAR-FORMATION HISTORY, AS TRACED BY
RADIO SOURCE EVOLUTION.}

\author{J.S. DUNLOP}
\institute{Institute for Astronomy, Department of Physics \&
Astronomy, University of Edinburgh, Royal Observatory, Edinburgh EH9 3HJ,
UK. }

\end{opening}

\runningtitle{COSMIC STAR-FORMATION \& RADIO SOURCE EVOLUTION}

\begin{document}

\begin{abstract}

{\small I briefly review our current knowledge of the cosmological
evolution of radio sources, and show that the redshift distributions of
new complete samples of radio sources confirm the existence of the
high-redshift decline in comoving number density (or `cutoff') beyond
$z \simeq 2.5$ first deduced by Dunlop \& Peacock (1990). 
Taken at face value these new data favour a luminosity
dependent `cutoff', in which the decline is least drastic for the most
luminous radio sources. I demonstrate, however, that regardless of the
precise form of radio source evolution, the evolution of
radio luminosity density is well determined, and appears uncannily
similar to the evolution of ultra-violet luminosity density from
star-forming galaxies in the universe ({\it e.g.} Madau 1997). 
I convert radio
luminosity density into an estimate of black hole fueling rate per Mpc$^3$,
and conclude that radio
source evolution is a good tracer of the star-formation history of the
Universe; at any epoch, for every 10$^7$ M$_{\odot}$ of material converted into
stars, approximately 1 M$_{\odot}$ appears to be 
consumed by radio-loud active galactic nuclei. If this is indeed true
at all epochs, then this would imply that star-formation activity in the
Universe peaked at $z \simeq 2 - 2.5$, and that the values of
star-formation rate density currently derived from Lyman-limit galaxies
at $z \simeq 2.8$ and $z \simeq 4$ are under-estimates by a factor
$\simeq 4$. Finally I briefly speculate 
as to why radio source evolution might trace global
star-formation activity as manifested primarily in the
disc/spiral population, whereas the hosts of the radio sources themselves
generally appear to be well-evolved elliptical galaxies.}

\end{abstract}

\section{Introduction}

The past year has seen the first meaningful attempts to determine the
global star-formation history of the Universe, using the combined
leverage provided by deep redshift surveys ({\it e.g.} the Canada France
Redshift Survey (CFRS); Lilly {\it et al.} 1995) reaching $z \simeq 1$, and the detection/non-detection
of Lyman limit galaxies at higher redshifts (in, for example, the Hubble Deep 
Field; Madau {\it et al.} 1996). 
The results ({\it e.g.} Madau 1997) indicate that star-formation
rate (and metal production) is $\simeq 10$ times greater at $z \simeq 1$
than in the local Universe, peaks somewhere around $z
\simeq 1-2$, and declines to values comparable with the present day at $z
\simeq 4$. The impact of dust obscuration on this result has yet 
to be fully assessed (Pettini, priv. comm.), 
as has the effect of combining samples selected 
in rather different ways. However, amid the excitement caused by this
revolution in optical cosmology, it seems to have been forgotten that
there already exists a class of source the evolution of which has been
well studied out to $z \simeq 4$, is derived from well-defined complete
samples, and is immune from the effects of dust.

Since the benefits of studying cosmological evolution at radio
wavelengths are so clear, an outsider might reasonably ask why so much
effort has recently been directed at determining the cosmological evolution
of ultraviolet light in our Universe. The answer, of course, is a general
(and not unreasonable) disbelief that the evolution of these rare and
bizarre fireworks -- powerful radio galaxies -- can tell us much (or indeed
anything) about the evolution of the `normal' galaxies of stars.
However, in recent years it has become clear that the level and form of
evolution displayed by powerful radio sources is not unique to either
this particular class of active galaxy or to the radio waveband. 
As has been discussed elsewhere ({\it e.g.} Dunlop 1994) at least out to
$z \simeq 2$ very similar evolution is found at optical wavelengths 
for optically-selected QSOs (Hewett {\it et al.} 1993) 
and at X-ray wavelengths for X-ray selected
quasars (Boyle {\it et al.} 1993). More intriguing still, however, is the suggestion that the
starburst galaxy population discovered by IRAS also displays similar
evolution (Rowan-Robinson {\it et al.} 1993), at least at low $z$.

This last comparison suggests that the form of evolution displayed by radio 
sources, as well as being applicable to AGN in general, might be 
of even wider relevance. In this brief article I have therefore 
explored the possibility that this evolution is truly universal,
by comparing the evolving radio luminosity density 
from powerful radio sources with recent estimates of the star-formation
history of the Universe. First, however, I reassess our current knowledge
of the high-redshift evolution of radio sources by comparing 
the predictions of the alternative high-redshift `cutoff' models of 
Dunlop \& Peacock (1990) with the actual redshift
distributions of two new complete samples; a `bright' (2 Jy) 
sample with complete redshift coverage, and the 
first `faint' sample (1 mJy) with spectroscopic redshifts or `reliable' 
estimates for all sources.

\section{The Cosmological Evolution of Radio Sources}

The advantages of completeness 
and immunity to dust offered by radio selection
can only be realised with reliable 
redshift information, which is hardest to achieve at high redshift.

\subsection{`Low'-Redshift Evolution}

Out to redshifts $z \simeq 2$ the evolution of powerful ($P_{2.7GHz} > 
10^{26} {\rm W Hz^{-1} sr^{-1}}$) radio sources is reasonably 
well-constrained, and in 1990 John Peacock and I showed that, given the  
existing complete-sample database, the 
evolution of both the flat-spectrum ($\alpha < 0.5$ where $f_{\nu}
\propto \nu^{-\alpha}$) and steep-spectrum ($\alpha > 0.5$) 
radio-source populations 
is at least consistent with pure luminosity evolution (PLE) 
with $P(z) \propto (1+z)^3$
(Dunlop \& Peacock 1990).

\subsection{High-Redshift Evolution: the Redshift CutOff Revisited}

The key component of the complete-sample database used by Dunlop \& Peacock 
(1990) to constrain the high-redshift evolution of the radio luminosity
function (RLF) was the Parkes 
Selected Regions (PSR), a sample of 178 sources with 
$S_{2.7GHz} > 100$ mJy over an area of 0.075 sr. Much of the evidence for a 
high-redshift decline in the steep-spectrum population then 
depended on 
$K$-band photometry to estimate the redshifts of the faintest galaxies in 
this sample. The reliability of this method has since been called into 
question (Eales {\it et al.} 1993) although interestingly a 
renewed spectroscopic campaign on the PSR is currently
revealing that virtually all the galaxies have true redshifts which 
are {\it smaller} than their $K-z$ estimates.
The revised redshift distribution for the PSR, and its 
implications for the high-redshift evolution of the RLF will be presented
elsewhere. However, a complete redshift distribution has now been  
obtained for a smaller and brighter complete sub-sample of 65 
radio galaxies from the 6C/B2 survey (Eales \& Rawlings 1996). 
Below I compare this   
redshift distribution with the predictions of the redshift
cutoff models of Dunlop \& Peacock (1990), and then consider 
the redshift distribution that John Peacock and I 
have recently derived for a much deeper 
sample from the Leiden Berkeley Deep Survey (LBDS), a sample 
of sufficient depth to resolve the issue of the redshift cutoff beyond
doubt.

\subsubsection{Alternative models for the redshift cutoff}

Although Dunlop \& Peacock (1990) concluded that the evidence for a 
high-redshift cutoff in the steep-spectrum luminosity function was strong, 
uncertainties in redshift estimation, combined with lack of coverage of 
the luminosity baseline, meant that we were unable to distinguish
between universal negative pure luminosity evolution at $z > 2.5$
(the PLE model) and an alternative model involving continuing positive
luminosity evolution combined with negative density evolution at 
high $z$ (the LDE
model). These two alternative models are illustrated in Fig. 1, the
principle difference between them being that in the LDE model the
strength of the decline in comoving number density is a function of radio
power.

\begin{figure} 
\vspace{42.4pc} 
\includegraphics{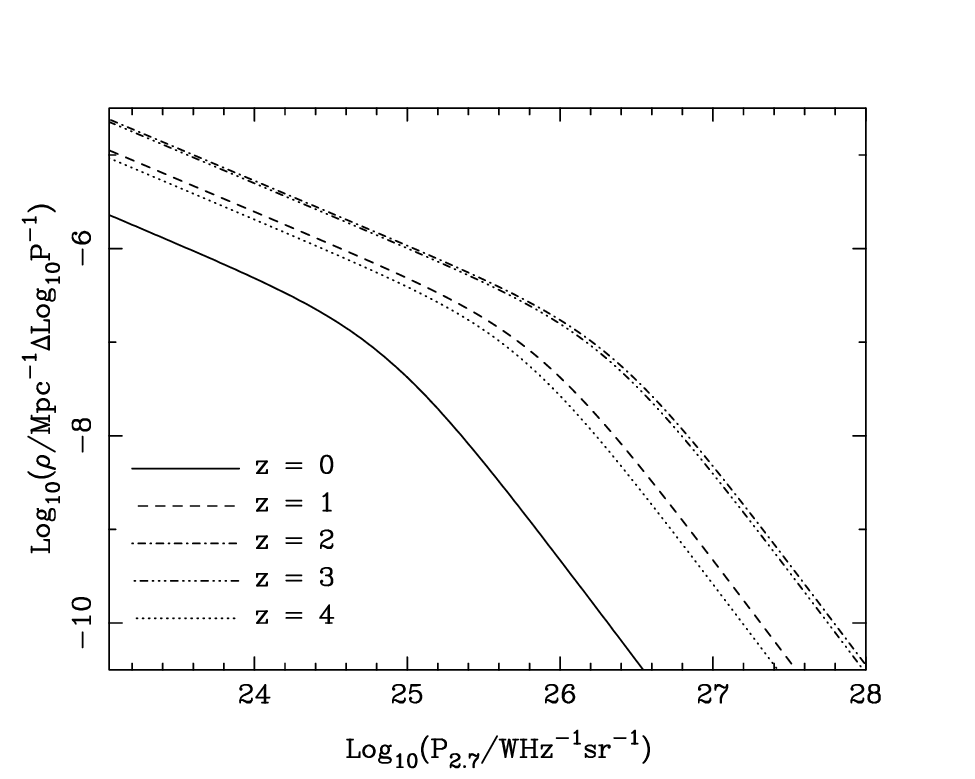} 
\includegraphics{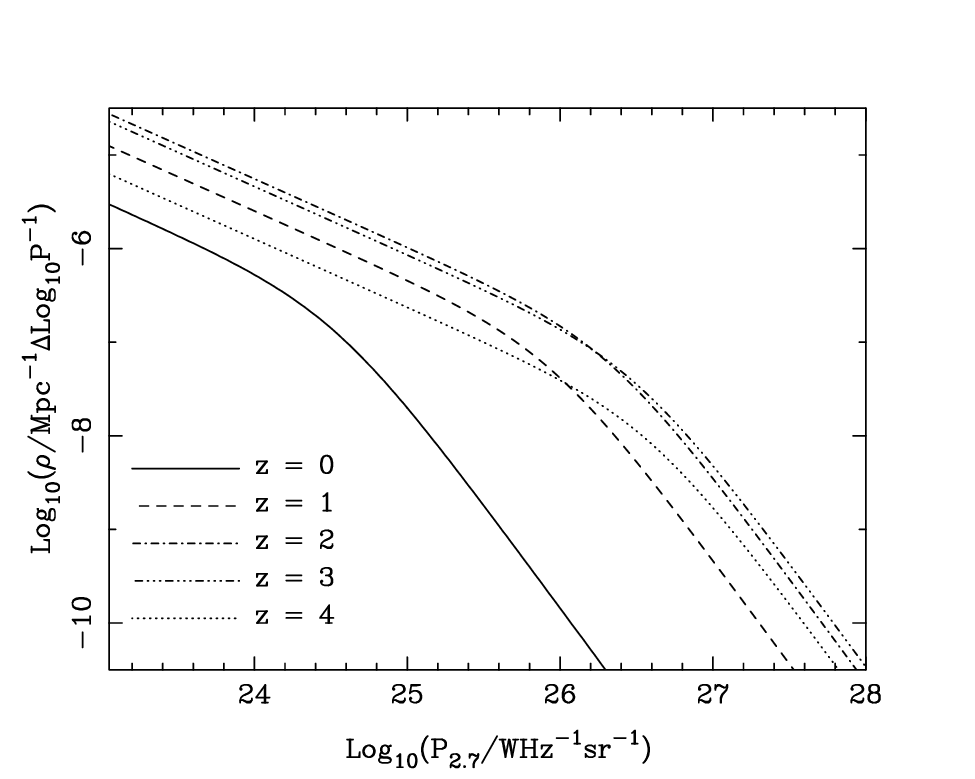} 
\caption[]{Two simple alternative models produced by Dunlop \& Peacock
(1990) to describe the high-redshift decline 
in the comoving number density of steep-spectrum radio sources. The upper
panel shows the PLE model 
in which the redshift cutoff is 
parameterized in terms of universal negative luminosity evolution at high
$z$. The lower panel shows the LDE model in which continuing positive
luminosity evolution is overcome by negative density evolution at all
powers for $z > 3$. In the latter model the apparent severity of the redshift 
cutoff 
becomes a function of radio luminosity, with the most luminous sources
suffering a less dramatic, but still significant decline 
in comoving number density.}
\end{figure}

\subsubsection{Comparison with the complete 6C sub-sample}

The complete 6C/B2 sub-sample consists of 65 sources detected at
151 MHz with ${\rm 2.2 Jy} < S_{151MHz} < {\rm 4 Jy}$ in 
0.1 sr of sky. Eales, Rawlings and
collaborators have recently completed the measurement of spectroscopic
redshifts for all the sources in this sample; 11 galaxies lie at $z > 2$ 
(Eales \& Rawlings 1996) and 
two of these lie at $z > 3$ (6C 1232$+$39;$z = 3.221$ and 
B2 0902$+$34;$z = 3.395$).
The cumulative redshift distribution for 
this sample for $z > 2$ is shown in Fig. 2a, where it is compared  
with the distributions predicted by the PLE and LDE
redshift cutoff models shown in Fig. 1, and also with the prediction of 
a `no-cutoff' model in which the form and normalization of the RLF is 
frozen for $z > 2$. The predicted number counts at 151
MHz have been 
produced from the models (which are defined at 2.7 GHz) 
by using the average spectral index displayed between 151 MHz
and 2.7 GHz by the 11 $z > 2$ sources 
in the sample ($\langle \alpha_{151MHz}^{2.7GHz} \rangle = 0.87$) to convert
the 151 MHz flux-density boundaries to their equivalent values at 2.7 GHz.
(${\rm 0.164 mJy} < S_{2.7GHz} < {\rm 0.298 mJy}$). 
It is clear that the redshift
distribution of this sample strongly supports the existence of the redshift
cutoff, but that the form of this cutoff seems better described by the LDE 
model than the PLE model. I note in passing that 
the recent discovery of 6C 0140$+$326 at $z = 4.41$ in a 
deeper 6C ($S_{151MHz} = 1 \rightarrow 2$
Jy) 0.1 sr sample (Rawlings {\it et al.} 1996) 
is also perfectly consistent with 
the LDE model of the high-redshift cutoff (see Dunlop 1996)

\begin{figure} 
  \vspace{42.4pc} 
\includegraphics{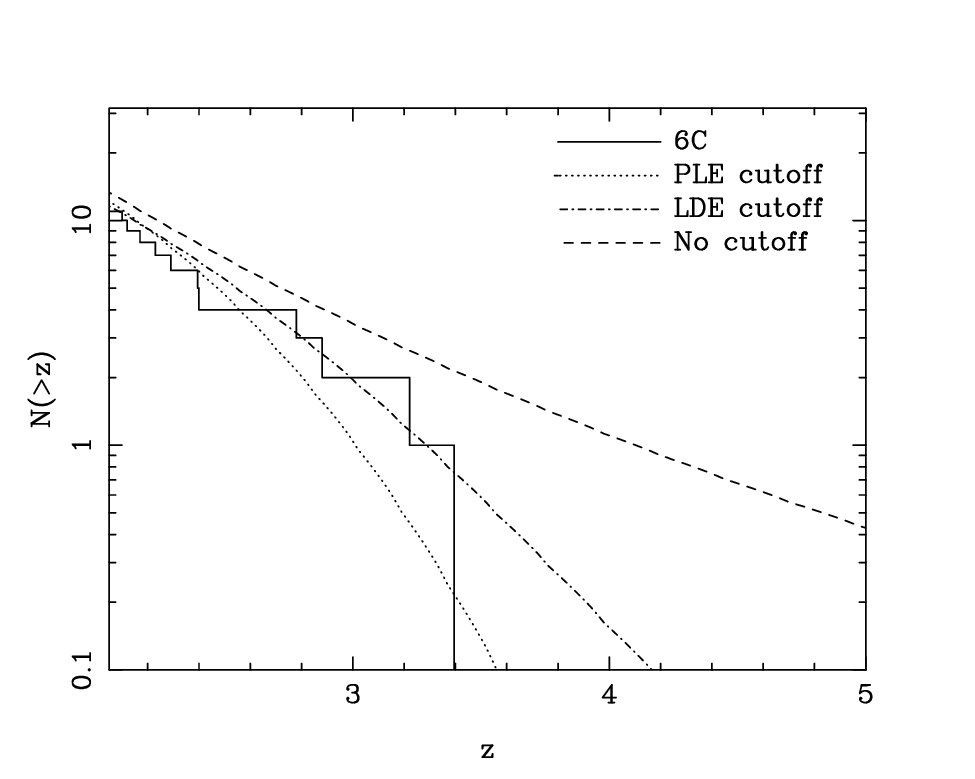} 
\includegraphics{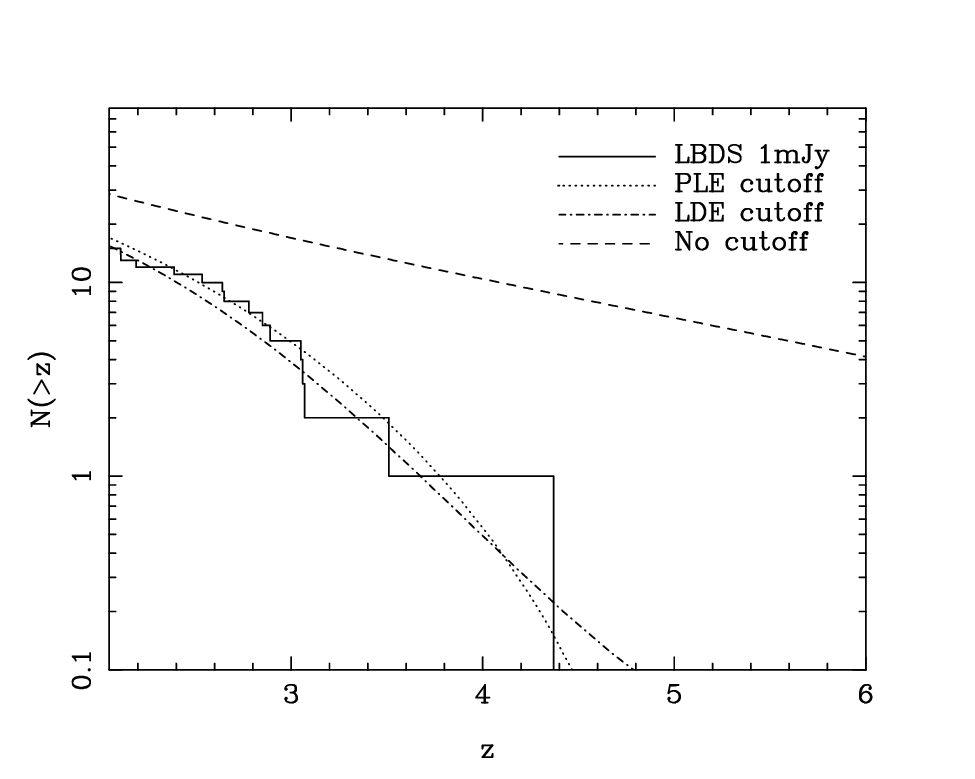} 
\caption[]{{\bf Upper Panel:} 
The observed high-redshift cumulative redshift distribution for
the complete 6C/B2 sub-sample, compared with the predictions of the PLE
and LDE cutoff models presented in Fig. 1, and with the predicted
redshift distribution for a model in which the radio luminosity function
is fixed at its $z \simeq 2$ form for all higher redshifts.
{\bf Lower Panel:} The high-redshift cumulative redshift distribution of the LBDS
($S_{1.4GHz} > 1$ mJy) sample, compared with the PLE, LDE and no-cutoff
model predictions assuming $\alpha_{1.4GHz}^{2.7GHz} \simeq 0.8$.}
\end{figure}

\subsubsection{Comparison with the new LBDS sample}

The 6C/B2 sub-sample considered above is of interest because of
its complete redshift information rather than because it can
really confirm or refute the existence of the redshift cutoff. 
Indeed it is a factor $\simeq 2$ {\it less} deep than the PSR, and thus at $z > 3$ can only be used to estimate the comoving
space density of objects brighter than $P_{2.7GHz} \simeq 10^{27} {\rm W
Hz^{-1} sr^{-1}}$ (which are intrinsically very rare). 
To unambiguously confirm or refute the existence of
the redshift cutoff really requires the study of a sample 
which is $\simeq 100$ times fainter, and is thus capable of
sampling the radio luminosity function down to $P_{2.7} \simeq
10^{24} {\rm W Hz^{-1}sr^{-1}}$ out to $z \simeq 4$ ({\it i.e.} below the break
luminosity at all redshifts -- see Fig. 1).

Accordingly, over the last few years we have been attempting 
to determine the redshift distribution of a statistically complete sample
of 77 galaxies with $S_{1.4GHz} > 1$ mJy selected from the LBDS (Windhorst {\it et al.}
1984a, 1984b, Kron {\it et al.} 1985). We now possess $g,r,i,J,H,K$ photometry
for the galaxies in this sample 
enabling us to estimate redshifts both from spectral fitting and
from a modified version of the $K-z$ diagram (Dunlop {\it et al.} 1995). Optical spectroscopy of a subset of sources 
with the Keck telescope shows this dual-pronged approach to
redshift estimation to be reliable, certainly out to $z \simeq
2.5$, principally because the starlight from these 
more moderate-luminosity radio galaxies is less contaminated by
strong emission lines or scattered AGN light than in the more
extreme high-$z$ objects found in brighter radio 
samples (Dunlop {\it et al.} 1996).

The resulting redshift distribution of this 1 mJy sample is compared with
that predicted by the PLE-cutoff, LDE-cutoff and no-cutoff models in
Fig. 2b, where the predicted redshift distributions have been produced
assuming $\alpha_{1.4GHz}^{2.7GHz} \simeq 0.8$. Comparison of the number count
predictions in this figure with those in Fig. 2a makes
clear the enormous power of this much deeper sample, despite the need to
resort to redshift estimation. In fact, to remove the cutoff,
10 of the 77 sources in this sample need to lie at $z > 4$, whereas 
our best estimate of the redshift distribution follows almost
exactly the predictions of the cutoff models.

\section{Universal Evolution}

The combination of the LBDS and 6C samples 
spans sufficient baseline in radio power to allow a first attempt 
at differentiating between
the PLE and LDE models, and it is clear that
Fig. 2 favours a luminosity-dependent cutoff
which is least drastic for the most luminous sources. Large
redshift surveys of bright radio quasars have confirmed that 
a very similar, perhaps also luminosity-dependent redshift cutoff 
is displayed by the
quasar RLF (Dunlop \& Peacock 1990; Shaver {\it et al.} 1996). 
Futhermore, it is now clear that the population of
optically-selected QSOs declines at high redshift with, 
arguably,  a comparable luminosity dependence 
(Warren, Hewett \& Osmer 1994; Kennefick
{\it et al.} 1995). 
The implication is that the similarity between
the evolving flat-spectrum RLF, steep-spectrum RLF and QSO OLF seen at $z < 2$
extends right out $z \simeq 4$, and that all powerful AGN suffer a similar
form of decline at $z > 2.5$. 

However, as is often the case in luminosity function studies, focussing
on uncertainty in the precise form of the evolving luminosity function
can cloud the fact that evolving {\it luminosity density} is rather 
robustly determined.
In Fig. 3 I plot the evolving luminosity-weighted integral
of the PLE and LDE RLFs. 
Both models yield essentially the same evolving 
luminosity density of radio emission out to $z \simeq 5$. In Fig. 3 the
evolving luminosity density of radio emission is presented in terms of an
evolving black hole fueling rate per Mpc$^3$ (left-hand axis; see figure 
caption for details) to enable ease of comparison with the evolving
star-formation rate per Mpc$^3$ deduced by Madau (1997) (data points and
right-hand axis in Fig. 3). The similarity between
the radio-based curves, and the ultraviolet-based data points in Fig. 3
appears too good to be a coincidence; the agreement is essentially perfect out
to $z \simeq 1$ where the star-formation rate is well determined, and at
$z > 1$ the curves are undoubtedly consistent with the limits derived
from the number density of Lyman-limit galaxies. Indeed, 
given the superior completeness of the radio surveys one might go so far
as to suggest that the curves shown in Fig. 3 provide the current `best
bet' as to the true star-formation density between 
$z \simeq 1$ and $z \simeq 5$. To bring the data-points at $z \simeq 3$
and $z \simeq 4$ into good agreement with the radio-based prediction
requires that the star-formation rate density as currently derived 
from the Lyman-limit galaxies is under-estimated by a factor $\simeq 3 - 4$
relative to the star-formation census provided at $z < 1$ by the CFRS.   
It will be interesting to see whether this transpires to be the case, or
whether star-formation activity really does peak at redshifts closer to
$z \simeq 1$.

At first sight it perhaps seems unlikely that cosmic 
star-formation activity, which at least at $z < 1$ occurs predominantly
in disc/irregular systems, should be traced by the evolution of powerful
radio sources which themselves are generally found 
in old giant elliptical galaxies
(even at $z \simeq 1.5$ at least some radio galaxies are $> 3$ Gyr old;
Dunlop {\it et al.} 1996, Dunlop 1997).
However, star-formation rate density presumably reflects the global rate of
gravitational accretion/condensation of gas at a given epoch, and it is
at least plausible that when such material falls into a massive galaxy 
containing a black hole, its mass is partly reprocessed as radio emission
rather than simply forming a disc of stars. Averaged over a large enough
volume, it might not be unreasonable to find that radio luminosity
density should reflect the global level of SFR at any given epoch.
It remains to be determined
whether the stars in the giant elliptical radio-source hosts 
are themselves formed in the tail of the distribution illustrated in
Fig. 3, or whether there exists a separate high-redshift peak of
star-formation corresponding to the (perhaps dust-enshrouded) formation
epoch of the most massive objects.

\begin{figure} 
  \vspace{21.2pc} 
\includegraphics{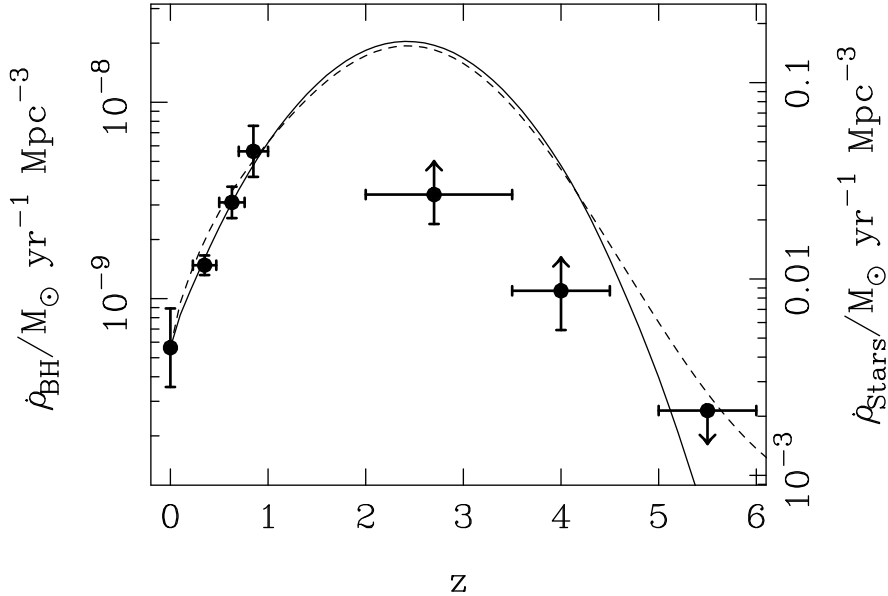} 
\caption[]{A comparison of the redshift dependence of 
the rate at which mass is consumed by black 
holes at the centre of giant elliptical galaxies (and turned into 
radio luminosity from AGN; curves 
and left-hand axis) with the rate at which mass is converted into stars 
per Mpc$^{3}$ (and turned into
UV-light from primarily disc/irregular galaxies; data points and 
right-hand axis). 
The solid and dashed curves
are the luminosity-weighted integrals of the PLE and LDE evolving radio
luminosity functions shown in Fig. 1, converted into black hole 
mass consumption rate per Mpc$^3$ assuming an efficiency of $\simeq 1$\%. 
The data-points
indicating the star-formation history of the Universe are taken from
Madau (1997), and are themselves derived from a low-redshift H$\alpha$
survey ($z \simeq 0$), the Canada France Redshift Survey ($z < 1$), and the 
number of colour-selected `U-dropout',
`B-dropout' and (lack of) `V-dropout' galaxies in the Hubble Deep Field.
The upward pointing arrows indicate the fact that the assessment of
star-formation rates based on Lyman-limit galaxies is liable to be
under-estimated due to the effects of dust.}
\end{figure} 

\end{document}